\documentclass[aps,superscriptaddress,showpacs]{revtex4}

\usepackage{graphicx}
\usepackage{amsmath,amssymb,amsfonts}
\usepackage{bm}
\usepackage{bibentry, natbib}
\usepackage[T1]{fontenc}
\usepackage[english]{babel}
\usepackage[latin1]{inputenc}
\usepackage{ifthen}
\usepackage{footnote}
\usepackage{subfigure}
\usepackage{setspace}
\usepackage{color}

\newcommand{\be}{\begin{equation}}
\newcommand{\ee}{\end{equation}}
\newcommand{\ba}{\begin{align}}
\newcommand{\ea}{\end{align}}
\newcommand{\sysb}{\left\{\begin{array}}
\newcommand{\syse}{\end{array}\right.}
\newcommand{\baa}{\begin{array}}
\newcommand{\eaa}{\end{array}}
\newcommand{\mal}{\mathcal}
\newcommand{\rmd}{{\rm{d}}}

\newcommand{\mand}{\quad\text{and}\quad}
\newcommand{\lan}{\left\langle}
\newcommand{\ran}{\right\rangle}
\newcommand{\vphi}{\varphi}
\newcommand{\eval}[1]{\left.\right|_{ #1 }}
\newcommand{\reff}[1]{(\ref{#1})}
\newcommand{\ce}{\rho}

\addto\captionsenglish{}

\begin{document}


\title{Collective non-equilibrium dynamics at surfaces and the spatio-temporal edge}
\author{Matteo Marcuzzi}
\author{Andrea Gambassi} 
\affiliation{SISSA -- International School for Advanced Studies and INFN, via Bonomea 265, 34136 Trieste, Italy}
\author{Michel Pleimling} 
\affiliation{Department of Physics, Virginia Tech, Blacksburg, Virginia, 24061-0435, USA}


\begin{abstract}

Symmetries represent a fundamental constraint for physical systems and relevant new phenomena often emerge as a consequence of their breaking. An important example is provided by space- and time-translational invariance in statistical systems, which hold at a coarse-grained scale in equilibrium and are broken by spatial and temporal boundaries, the former being implemented by surfaces --- unavoidable in real samples --- the latter by some initial condition for the dynamics which causes a non-equilibrium evolution.
While the separate effects of these two boundaries are well understood, we demonstrate here that additional, unexpected features arise upon approaching the effective edge formed by their intersection.
For this purpose, we focus on the classical semi-infinite Ising model with spin-flip dynamics evolving out of equilibrium at its critical point.
Considering both subcritical and critical values of the coupling among surface spins, we present numerical evidence of a scaling regime with universal features which emerges upon approaching the spatio-temporal edge and we rationalise these findings within a field-theoretical approach.

\end{abstract}


\maketitle

\emph{Introduction --} Thanks to the advances in miniaturisation of the past decades, devices have reached a size at which boundary effects cannot be neglected; in parallel, the physics of surfaces and interfaces has attracted an increasing interest, in particular concerning non-equilibrium dynamical processes, as many applications involve sudden changes in control parameters. Typically, describing these processes requires knowledge of the many microscopic details that vary widely from system to system. However, in suitable circumstances, collective phenomena emerge which make only few coarse-grained, mesoscopic properties relevant for the dynamics: in fact, it is now well-established, both theoretically and experimentally, that the behaviour of a statistical system close to a continuous phase transition can be characterised by few quantities, such as exponents and scaling functions, which depend only on the range and symmetries of the underlying interaction and on the dimensionality of the space. All the microscopically different systems sharing these same gross features form the so-called universality class of the transition. Within each of these classes, the various thermodynamic and structural properties typically show, in the neighbourhood of the critical point, leading algebraic behaviours characterised by common exponents, which constitute the hallmark of the transition. In turn, upon approaching it, the relevant contribution to the various thermodynamic quantities is effectively determined by the fluctuations of the so-called order parameter $\vphi$ (\textit{e.g.}, the local magnetisation for an Ising ferromagnet).

The emergence of universality is currently understood within the framework of the renormalisation group (RG) \cite{Wilson, Zinn-Justin}, which is among the most important theoretical achievements of the past forty years. RG transformations highlight scale invariance, a feature that emerges upon approaching a continuous phase transition, as the only mesoscopic length-scale present --- \textit{i.e.}, the correlation length $\xi$ of the fluctuations of the order parameter --- ideally diverges; because of it, one can focus on effective, coarse-grained descriptions and models which retain only the gross features of the original system and which are effective at space and time scales larger than the microscopic ones. Originally devised for describing the local behaviour of finite samples far from the boundaries (\textit{i.e.}, in their bulk), this theoretical tool has been subsequently extended in order to account for the presence of surfaces (\textit{e.g.}, see Refs.~\cite{Binder, Diehl}). Their introduction alters locally the properties of the system and breaks translational invariance: within the RG approach, a new set of relevant parameters has to be introduced in order to account for the gross features of the boundary. As a result, novel singularities might emerge upon approaching the surface, which split the original universality class in surface subclasses characterised by a set of boundary exponents and scaling functions associated, \textit{e.g.}, with the algebraic behaviour of the $\vphi$ correlation functions in its proximity \cite{Binder,Diehl}. In general, these exponents cannot be inferred from the bulk ones.
A number of analytical \cite{Special,Ordinary,Diehl2}, numerical \cite{Num4,Num5} and experimental (see, \textit{e.g.}, Ref.~\cite{ExpRef}) studies investigated primarily semi-infinite and film geometries, whereas wedges, edges \cite{Cardy,Pleimling3}, as well as curved and irregular surfaces \cite{Diehl,Pleimling2} were studied to a lesser extent. Universal features emerge also in the dynamic behaviour at equilibrium (both in infinite and finite systems) \cite{NonEqFinite}, and the universality class is further split depending on the gross features of the dynamics, such as the possible global conservation of the order parameter \cite{HH,EqSurface}. 

A sudden quench of the system's temperature breaks time-translational invariance --- a characteristic symmetry of stationary states --- and, analogously to the case of spatial surfaces, effectively introduces a sharp temporal boundary for the evolution and causes non-equilibrium behaviours to emerge \cite{AgingDyn, Janssen}. In particular, the memory of the initial state naturally affects the dynamics close to this boundary and, in certain circumstances, it is also responsible for the later occurrence of ageing phenomena \cite{Janssen, AgingDyn, Gambassi}. Within this picture, the dynamics at "short" times becomes akin to the behaviour of static quantities at short distances from a spatial surface \cite{EqSurface} and therefore RG techniques can be applied in order to study the short-time and ageing behaviours and their emerging universal features \cite{Janssen, Gambassi}. 
In order to investigate the subtle interplay between the breaking of space- and time-translational invariance we study here the semi-infinite Ising ferromagnet quenched from a disordered state to its critical temperature at time $t=0$. This relatively simple, but paradigmatic case can be used to verify whether effects beyond those resulting from each separate breaking \cite{Ritschel} emerge. We use both numerical and analytical methods to show that, unexpectedly but similarly to the case of a spatial wedge \cite{Cardy}, a so-far undetected power-law behaviour described by a critical exponent $\theta_E$ emerges upon approaching the effective edge formed by the intersection of the spatial and temporal boundaries.

\emph{Scaling behaviour --} Analogously to the Ginzburg-Landau (GL) theory of phase transitions, the collective dynamical properties of Ising ferromagnets can be studied in terms of a continuum field theory for the order parameter $\vphi(\vec{x},t)$ (which represents here the coarse-grained local magnetisation), the thermal stochastic dynamics of which is captured by a Langevin equation \cite{HH}
\be 
\frac{\partial\vphi}{\partial t} = - \Omega \frac{\delta \mal{H}}{\delta\vphi} + \eta ,
\label{eq:Langevin}
\ee
where $\eta(\vec{x},t)$ is a white Gaussian noise with zero mean and variance $\left\langle  \eta(\vec{x}, t)\eta(\vec{y}, s)\right\rangle = 2\Omega k_B T \delta\left( \vec{x}-\vec{y} \right) \delta\left( t-s \right)$ and $\mal{H}$ is the static GL free-energy \cite{Zinn-Justin}
\be 
\mal{H}[\vphi] = \!\!\!\!\!\! \int\limits_{\left\{ x_\perp \geq 0 \right\} }\!\!\!\! \rmd^{d} x  \left[  \frac{(\vec{\nabla}\vphi)^2}{2} + \frac{r}{2}\vphi^2 + \frac{g}{4!}\vphi^4 +  \delta(x_\perp)\frac{c_0}{2} \vphi^2 \right] .
\label{eq:GL}
\ee
Here $x_\perp$ stands for the coordinate normal to the surface located at $x_\perp=0$, $d$ is the space dimensionality and, denoting by $T$ the temperature and by $T_c$ its critical value, $r \propto T - T_c$. Whereas $g>0$ characterises the interaction strength in the bulk, the so-called surface enhancement $c_0$ is related to the difference between the values $J_b$ and $J_s$ of the coupling constant among the spins in the bulk and within the surface, respectively. Depending on its value, one distinguishes three cases \cite{Binder, Diehl}: (i) $c_0 \to +\infty$ (ordinary transition) corresponds to $J_s \ll J_b$ which implies that the surface ordering is induced by the bulk one; (ii) $c_0 \to -\infty$ (extraordinary transition) corresponds to the case $J_s \gg J_b$, in which the surface acquires a non-vanishing magnetisation independently of the bulk; these two instances are separated by a critical value (iii) $c_0 = c_{sp}$ (special transition), which corresponds to a certain ratio $J_s/ J_b$ ($\simeq 1.5$ for the three-dimensional Ising model \cite{Hasenbusch}). The last two cases are present only for $d\geq 3$. Hereafter $\lan\cdot\ran$ indicates the average over the possible realisations of the thermal noise $\eta$ and of the initial state; for convenience, we set the scales of time and $T$ such that $\Omega = k_B T_c = 1$ and fix the system at its bulk critical point $T = T_c$.
The initial condition is chosen to be an effective high-temperature state both in the bulk and at the surface, characterised by local correlations $\langle \vphi(\vec{x}, t=0)  \vphi (\vec{y}, t=0) \rangle = \tau_0^{-1} \delta (\vec{x}-\vec{y})$ with $\tau_0$ much larger than any other scale, \textit{i.e.}, we will actually consider the limit $\tau_0 \to +\infty$.
In the following we shall concentrate on the ordinary and special transitions. The corresponding average order parameter $\langle \vphi (t) \rangle$ was studied in Ref.~\cite{Ritschel}, where a power-counting argument was proposed in order to argue that no new field-theoretical divergences arise at the spatio-temporal edge ($t=0,\,x_\perp=0$). Accordingly, the exponents of the associated algebraic singularities should be a combination of those characterising the static bulk ($\beta$, $\nu$) and surface ($\beta_1$) behaviours and the bulk equilibrium ($z$) and non-equilibrium ($\theta$) dynamics. For the total magnetisation $\Phi(x_\perp,t)$ of the plane at a distance $x_\perp$ from the surface (\textit{i.e.}, the spatial integral of $\vphi(\vec{x},t)$ along the $d-1$ directions parallel to it) the two-point and two-time correlation function
\be
C(x,t;y,s) = \lan \Phi(x,t) \Phi(y,s)  \ran
\label{eq:twopoint}
\ee
is expected to scale as \cite{Sengupta,Gambassi}
\begin{align}
C(x,t;y,s) = &\ (t-s)^{a}\left( \frac{s}{t} \right)^{1-\theta} \left( \frac{ A^2 xy}{(t-s)^{2/z}} \right)^{(\beta_1 -\beta)/\nu}\nonumber\\
&\times F_C \left( \frac{(Ax)^z}{t-s},\frac{(Ay)^z}{t-s}, \frac{s}{t} \right),
\label{eq:C}
\end{align}
where $A$ is a non-universal constant which makes the scaling variables dimensionless and is of order unity for a suitable choice of time and space units. $F_C$ is a scaling function (depending, inter alia, on the surface transition) with finite $F_C(0,0,0) \neq 0$ and such that the usual equilibrium scaling is recovered in the bulk and at the surface when $t \to s$. For the $d=3$ Ising class, $a = (d-1-2\beta / \nu)/z \simeq 0.96$ \cite{Vicari,Pleimling}, $\nu \simeq 0.63$ \cite{Vicari} is the exponent of the correlation length, $\beta \simeq 0.33$ \cite{Vicari} and $\beta_{1,ord} \simeq 0.80$ (or $\beta_{1,sp} \simeq 0.24$) \cite{Pleimling2} are the order parameter exponents in the bulk and at the surface, $z \simeq 2.04$ \cite{Pleimling} is the bulk dynamic critical exponent, whereas $\theta \simeq 0.15$ \cite{Pleimling} governs the short-time behaviour in the bulk. Note that in eq.~\reff{eq:C} $\beta_1$ is the only exponent that distinguishes the ordinary from the special transition. We now define what we refer to as the "edge regime" (E): $x$, $y$ and $t$ are fixed such that $y^z \ll t-s$, while $s$ varies within the range $y^z \ll s \ll t$; correspondingly, eq.~\reff{eq:C} yields $C(\ldots,s) \propto s^{1-\theta}$, where the proportionality constant depends, inter alia, on $t \gg s$, $x^z/t$ and $y^z/t \ll 1$. Note that, according to eq.~\reff{eq:C}, the assumption $ y^z \ll s$ is not necessary for observing this power law; however, it becomes important for the extension discussed further below. Conversely, we name "short-time" (ST) the regime in which $s$ is much smaller than any other scale (which can be obtained from the edge regime by moving $s$ to the domain $s \ll y^z$).
In summary, eq.~\reff{eq:C} predicts that the power-law behaviour of $C$ as a function of $s$ in the edge regime is the same both for the special and the ordinary case.
\begin{figure}[ht]
\includegraphics[width=0.50\columnwidth]{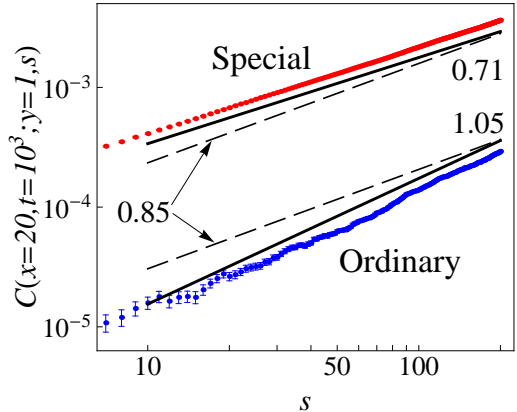}
\caption{(Color online) Time dependence of the correlation function $C(x,t;y,s)$ of the plane magnetisation of a three-dimensional Ising model evolving with Glauber dynamics after a quench from a completely disordered state to the bulk critical temperature. The bulk-surface correlation function $C$ with $x = H/2 = 20$ and $y=1$ is plotted as a function of $s$ for $t = 10^3$ sweeps (corresponding to the "edge regime" discussed in the main text). Data points result from averaging over $5\times 10^5$ and $3 \times 10^6$ independent runs for the special (red upper dots) and ordinary (blue lower dots) case, respectively. The black solid lines correspond to power laws with the exponents reported to the right; the parallel dashed lines, instead, indicate, for comparison, the power law with exponent $0.85$ observed within the short-time regime (see fig.~\ref{fig:Ising2}) and predicted by eq.~\reff{eq:C}. The data points with $s < 6$ have been disregarded as they might be affected by microscopic effects.}
\label{fig:Ising1}
\end{figure}

\emph{Numerical analysis --} Using the standard setup for studying surface criticality \cite{Pleimling2}, we simulate the Ising model on a three-dimensional cubic lattice made up of $H = 40$ consecutive planes with $ 60\times 60$ spins per plane. Within each of these planes, periodic conditions are imposed at the boundaries. The coupling constant between two surface spins is $J_s$, while in the rest of the lattice it is set to $J_b = 1$. The ordinary case is realised for $0 \leq J_s < 1.5$, whereas the special transition is known to occur at $J_s \simeq 1.5$ \cite{Hasenbusch}. The system is prepared at $t=0 $ in a completely disordered state corresponding to infinite temperature $T$ and it subsequently evolves at its bulk critical value $T_c = 4.5115 J_b/ k_B$ with Glauber dynamics. One time step corresponds to one sweep where on average every spin of the lattice has been updated once; as we are interested in the behaviour close to the temporal surface, only rather short simulation times are needed and therefore we can restrict ourselves to rather small system sizes. In order to rule out the influence of finite-size effects we also analysed some larger lattices. For the purpose of investigating the edge regime we study the two-point correlation function $C(x,t;y,s)$ defined in eq.~\reff{eq:twopoint}; here, $x=1$ corresponds to the surface, whereas $x = H/2$ to the midplane of the lattice. The statistical average is taken over a large number of independent runs with different realisations of the thermal noise and of the initial state. In fig.~\ref{fig:Ising1} we show the bulk-surface correlator $C(x = H/2, t; y=1,s)$ for the ordinary case $J_s = 1$ and the special one $J_s = 1.5$, where we fix $t = 10^3$ and vary $s$ within a range that, under the assumption $A \simeq 1$, corresponds to the edge regime (we verified that different values of $t$ give the same result). We observe that $C(\ldots,s) \sim s^{\ce}$ increases algebraically, but with an exponent $\ce_{sp} = 0.71(2)$ for the special case and $\ce_{ord} = 1.05(2)$ for the ordinary one. This unexpected result $\ce_{sp} \neq \ce_{ord}$ clearly shows that different surface universality classes display a different behaviour at the spatio-temporal edge, while eq.~\reff{eq:C} requires $\ce_{sp} = \ce_{ord} = 1 - \theta \simeq 0.85$ \cite{Pleimling}.

\emph{Field-theoretical results --} In order to rationalise these findings, we carried out a field-theoretical perturbative calculation of $C$ up to the first order in $\epsilon = 4-d$ ($d = 3$ in the simulations); beyond correctly reproducing the previously-known results concerning separately each of the two boundaries $x_\perp = 0$, $t=0$, this approach highlights indeed the emergence of an additional dimensional pole $\propto \epsilon^{-1}$ \cite{Zinn-Justin} when the coordinates of the correlation function \reff{eq:twopoint} are fixed at the spatio-temporal edge $y=s=0$, analogously to what happens for the static critical behaviour in a wedge \cite{Cardy}.
%
\begin{figure}[ht]
\includegraphics[width=0.50\columnwidth]{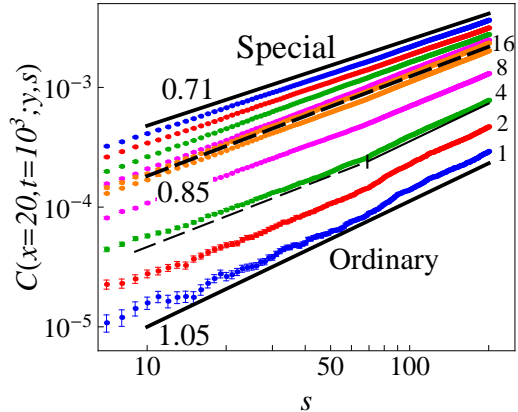}\\[0.3cm]
\caption{(Color online) Time dependence of the correlation function $C(x=20,t=10^3;y,s)$ for the special and ordinary transitions and various values of the position $y$. For the $5$ lowermost (uppermost) curves, corresponding to the ordinary (special) transition, $y$ increases from bottom to top (top to bottom) and takes the sequence of values reported for the ordinary case to the right. This plot highlights the crossover from the "edge" behaviour (thick solid lines), observed for $y^z \ll s$ in accordance with fig.~\ref{fig:Ising1}, to the "short-time" behaviour for $y^z \gg s$, indicated by the thick dashed line, which is common to the two instances. For the ordinary case and $y=4$ this crossover can be observed as a function of $s$, as highlighted by the thin solid and dashed lines that meet at $s \simeq (Ay)^z$, which yields the indicative estimate $A\simeq 2$. The details of the simulations are the same as for fig.~\ref{fig:Ising1}. }
\label{fig:Ising2}
\end{figure}
%
This pole is actually related to a dangerously irrelevant edge operator $\vphi_E$, which appears as the most relevant field in the short-distance expansion \cite{Diehl} of the order parameter $\vphi$ close to the edge; in the present case, due to the properties of the boundaries, $\vphi_E$ can be identified with $\partial_t \vphi(t,x_\perp)\eval{t=x_\perp=0}$ in the special case and with $\partial_t \partial_{x_\perp} \vphi(t,x_\perp) \eval{t=x_\perp=0}$ in the ordinary one \cite{Diehl,Janssen}. In order to account for its presence, one has to introduce a "radial" coordinate $(Ay)^z + s $, with $A$ fixed as in eq.~\reff{eq:C}, which controls the distance from the edge. Because of $\vphi_E$, the scaling form \reff{eq:C} acquires a new algebraic term, which can be made explicit by the substitution
\begin{align}
F_C\left( \ldots \right) \to \left(\frac{(Ay)^z + s}{t-s}  \right)^{-\theta_E}  \widetilde{F}_C \left( \ldots \right),
\label{eq:C2}
\end{align}
where now $\widetilde{F}_C(0,0,0) \neq 0$ is finite. While, as expected, this new scaling form reproduces the results of eq.~\reff{eq:C} when approaching either the spatial (\textit{e.g.}, fixed $s$ and $y \to 0$) or the temporal surface (fixed $y$ and $s \to 0$), the additional term in eq.~\reff{eq:C2} affects the behaviour of the correlation function in the edge regime as $C(\ldots,s) \sim s^{1-\theta - \theta_E}$, which yields $\ce = 1 - \theta - \theta_E$. The critical exponent $\theta_E$ encodes the anomalous dimension of $\vphi_E$ and, up to $O(\epsilon)$, turns out to be \cite{Marcuzzi}
\be
\theta_{E,sp} = \left(\sqrt{3} -1  \right)\frac{\epsilon}{12}  \mand \theta_{E,ord} = -\left(1 - \frac{1}{\sqrt{3}}     \right)\frac{\epsilon}{12} 
\label{eq:thE}
\ee
for the special and ordinary case, respectively. Table \ref{tab:comparison} presents the comparison between the numerical estimates of $\ce$ from fig.~\ref{fig:Ising1} and the corresponding analytic expressions resulting from Eqs.~\reff{eq:C} and \reff{eq:C2}, with (up to $O(\epsilon)$) $\theta = \epsilon / 12$ \cite{Janssen} and $\theta_E$ as in eq.~\reff{eq:thE}, specialised to the three-dimensional case $\epsilon =1$.
%
\begin{table}[ht]
\caption{Comparison between the numerical and  analytical  estimates of the exponent $\ce$ which controls $C(\ldots,s) \sim s^\ce$ in the edge (E) and the short-time (ST) regimes according to the scaling forms Eqs.~\reff{eq:C} and \reff{eq:C2}.}
\begin{ruledtabular}
\begin{tabular}{cccc}
  & Monte Carlo & \multicolumn{2}{c}{Analytical ($\epsilon =1$)} \\
  & Figs.~\ref{fig:Ising1} and \ref{fig:Ising2} & eq.~\reff{eq:C} & eq.~\reff{eq:C2}  \\
\hline
 E, sp & $0.71(2)$ & $0.917$ & $0.856$  \\
 E, ord & $1.05(2)$ & $0.917$ & $0.952$  \\
 ST, sp \& ord & $0.85(2)$ & $0.917$ & $0.917$  \\
\end{tabular}
\end{ruledtabular}
\label{tab:comparison}
\end{table}
%
Even if lacking quantitative accuracy, the additional term introduced in eq.~\reff{eq:C2} correctly captures the qualitative behaviour as a function of $s$ of the correlation function $C(x,t;y,s)$ near the spatio-temporal edge: in the case of the ordinary and special transition $C(\ldots,s)$ grows respectively faster and slower than it does in the short-time regime. Figure \ref{fig:Ising2} presents a study of the crossover between the edge regime of fig.~\ref{fig:Ising1} and the short-time regime, which occurs upon increasing $y$ above the scale set by $s^{1/z}/ A$. This crossover is properly captured by the scaling function in eq.~\reff{eq:C2} because the additional multiplicative factor becomes independent of $s$ for $s \ll t$, so that $C(\ldots,s) \sim s^{1-\theta}$, as predicted also by eq.~\reff{eq:C}. Conversely, for a fixed $y \neq 1$, the crossover between the short-time and the edge regime occurs upon increasing $s$ above $(Ay)^z$. With the present data, we could find clear evidence of this crossover for the ordinary case with $y=4$ (see fig.~\ref{fig:Ising2}). Note that the edge regime is not the only one affected by the aforementioned factor; however, we focused on it in order to test a qualitative difference between the predictions of eq.~\reff{eq:C} and \reff{eq:C2}, which does not hinge on a difficult, quantitatively accurate analytic determination of $\theta_E$.

\emph{Conclusions --} By studying the non-equilibrium dynamics of the Ising model in proximity to a surface, we have identified a regime 
in which the spatio-temporal edge, defined as the intersection between the spatial and temporal boundaries, affects the scaling behaviour of the correlation function; we rationalised and qualitatively accounted for the numerical evidence within a perturbative, field-theoretical framework. The consistency of these approaches, which concern two completely different systems at the microscopic level, \textit{i.e.}, a discrete spin model on a lattice and an interacting field theory in the continuum, provides additional support to the universality of the edge behaviour, which could be expected as a consequence of scale invariance at the critical point. This actually indicates that the scaling near the edge investigated here depends only on the gross features (symmetry, dimensionality and short-range interaction) which are shared by both models. Analogous results have been obtained for the linear response function and within the vector generalisation of the present model \cite{Marcuzzi}.
It would be desirable to extend the present investigation to different static, dynamic and surface universality classes of experimental relevance. While the first experiments probing the static surface and equilibrium dynamical bulk properties of materials date back to the 1970s \cite{HH, Exp1}, techniques with the sufficient accuracy to study bulk non-equilibrium behaviours have been available only since the 1990s \cite{Exp3}. Surface dynamics in condensed matter systems, instead, have not been observed until recently, though with a different purpose and not in the critical regime \cite{ExpSDynamics}. Extending recent investigations of ageing phenomena in liquid crystals \cite{Ciliberto} to the proximity of surfaces might provide a viable alternative for the experimental test of the present predictions in systems undergoing an Ising transition.

\emph{Acknowledgements --} M.M.~and A.G.~are grateful to J.~Cardy and S.~Majumdar for useful discussions. A.G.~is supported by MIUR within the program ``Incentivazione alla mobilit\'a di studiosi stranieri e italiani residenti all'estero''. M.P. acknowledges the support by the US National Science Foundation through DMR-0904999.

\end{document}